\documentclass{article}
\usepackage{spconf,amsmath,graphicx}
\usepackage{graphicx}
\usepackage{epstopdf}
\usepackage{booktabs}
\usepackage{calligra}
\usepackage{bm}
\usepackage{multirow}
\title{SCNet: Sparse Compression Network for Music Source Separation}
\name{\begin{tabular}{c}
Weinan Tong$^{1,2, \ddagger}$\thanks{$\ddagger$Work performed while interning at Skywork AI PTE.LTD.}, 
Jiaxu Zhu$^{1,\dagger}$\thanks{$\dagger$ Equal contribution.  *Corresponding author.}, 
Jun Chen$^{1}$ \\
Shiyin Kang$^{2,*}$, Tao Jiang$^{2}$, Yang Li$^{2}$, Zhiyong Wu$^{1,3,*}$, Helen Meng$^{4}$
\end{tabular}}

\address{
$^{1}$ Shenzhen International Graduate School, Tsinghua University, Shenzhen, China\\
 $^{2}$ Skywork AI PTE. LTD. $^{3}$ Peng Cheng Lab, Shenzhen, China\\
$^{4}$ The Chinese University of Hong Kong, Hong Kong SAR, China\\
\small{
        \{twn21, zhu-jx21, y-chen21\}$@$mails.tsinghua.edu.cn, 
        zywu$@$sz.tsinghua.edu.cn,
        shiyin.kang@kunlun-inc.com
    }
}
\begin{document}
\ninept

\maketitle

\begin{abstract}
Deep learning-based methods have made significant achievements in music source separation. However, obtaining good results while maintaining a low model complexity remains challenging in super wide-band music source separation. 
% The previous work either didn't consider the difference of subbands or 
% didn't handle information loss problem well 
% %didn't deal well with the information loss problem
% in generating subband features. 
Previous works either overlook the differences in subbands or inadequately address the problem of information loss when generating subband features.
In this paper, we propose SCNet, a novel frequency-domain network to explicitly split the spectrogram of the mixture into several subbands and introduce a sparsity-based encoder to model different frequency bands. We use a higher compression ratio 
on subbands with less information to improve the information density
and focus on modeling 
subbands with more information.
In this way, the separation performance can be significantly improved using lower computational consumption. Experiment results %also 
show that the proposed model achieves a signal to distortion ratio (SDR) of 9.0 dB on the MUSDB18-HQ dataset without using extra data, which outperforms 
state-of-the-art methods.
Specifically, SCNet's CPU inference time is only 48\% of HT Demucs, one of the previous state-of-the-art models.
\end{abstract}
\section{Introduction}
\label{sec:intro}

Music source separation (MSS) aims to separate the pure source signal and accompaniment from the mixed signal, such as separating human voice and accompaniment from singing. 
% Many of the music source separation models are motivated by existing system architectures like speech separation \cite{luo2019conv, luo2017deep}. Although directly using models in speech separation has proven effective in terms of the separation performance, this simple migration is difficult to achieve the best results without considering the difference between singing voice and speech in terms of the fundamental frequency, loudness, and formants \cite{livingstone2013acoustic}. 
%In recent years, many deep learning approaches have been proposed to solve the MSS problem.
Many prevailing MSS methodologies predominantly explore the utilization of the frequency-domain \cite{uhlich2017improving, chandna2017monoaural, hennequin2020spleeter, liu2021cws} or a fusion of time-domain and frequency-domain \cite{kim2021kuielab, defossez2021hybrid, rouard2023hybrid}. 
%But they often neglect the variances across musically distinct sources and the unique attributes inherent to different frequency bands.
While these approaches have yielded impressive results, they often overlook the disparities between musical signals across various frequency bands, which subsequently becomes a performance bottleneck.
BSRNN \cite{luo2023music} explicitly splits the spectrogram of the mixture into subbands and performs interleaved band-level and sequence-level modeling to get the state-of-the-art separation performance. However, the method of splitting subbands in BSRNN is very complicated, and it is necessary to try different schemes to get the best results. Even more, the band-splitting scheme needs to be further modified according to the sound of specific instruments, which further increases the manual design operation.

Moreover, music signals are typically recorded at a higher sample rate (e.g., 44.1k Hz) \cite{MUSDB18HQ}, which offers great details and enables neural network models to learn richer audio representations and achieve better separation performance. This, however, comes at the cost of linearly growing computational complexity, making them less deployable for resource-constrained applications. The simplest solution to this challenge is to downsample the audio to a lower sample rate. Nevertheless, this will drop the fine details captured from the high-sample rate sensor, and the omitted details could limit the model's performance \cite{rouard2023hybrid}. Dropping details uniformly at all positions is clearly suboptimal, as not all frequency parts are equally informative. Within a spectrogram, the lower-frequency part containing more detailed features is more critical than the higher-frequency part. Inspired by \cite{chen2023sparsevit}, sparse, high-resolution features are far more informative than dense, low-resolution ones. A very natural idea is to skip computations for less-important parts and focus on more-important details.

Therefore, in this paper, we propose an innovative frequency-domain network architecture called SCNet, which consists of an audio encoder, a separation network based on dual-path RNN \cite{luo2020dual}, and an audio decoder. Inspired by \cite{li2022efficient}, the proposed architecture of the encoder and decoder contains top-down, bottom-up, and skip connections, mirroring the brain's hierarchical processing of sensory information \cite{rouard2023hybrid, defossez2019demucs, stoller2018wave}. 
% The model takes a stereo mixture as input and outputs a stereo estimation for each source.
Considering the disparities among frequency subbands, we also employ different processing for them. Specifically, different from the intricate subband-splitting method used by BSRNN \cite{luo2023music}, we simply divide the spectrogram into three distinct frequency subbands: low, medium, and high, which dramatically reduces the complexity of splitting subbands. Moreover, unlike BSRNN, which has to set different parameters for different musical sources separately, we only need unified model parameters to separate multiple music sources. Furthermore, in order to make full use of high sampling rate music information to improve the separation performance and reduce the computational complexity of the model, we use different compression ratios and modeling operations for different frequency subbands. Recognizing that the low-frequency region is information-rich, while the mid- and high-frequency bands are comparatively less, we use a larger compression ratio for the mid-and high-frequency subbands to increase the information density. In addition, since the information in the low-frequency subband is more complex, we use relatively more parameters to focus on modeling it. In this way, the separation performance can be significantly improved using lower computational consumption. Experiment results show that the proposed model trained on the MUSDB18-HQ \cite{MUSDB18HQ} dataset substantially outperforms the state-of-the-art model with 9.0 dB SDR.  And the CPU inference time is only 48\% of HT Demucs.

\begin{figure*}[htb] 
\vspace{-0.1cm}
\center{\includegraphics[width=0.8\linewidth]  {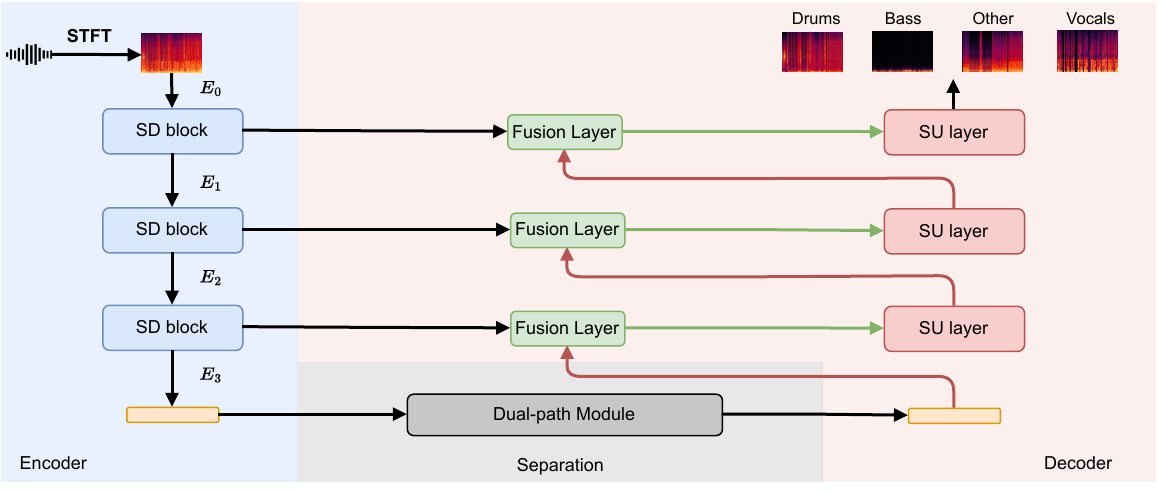}} 
\caption{\label{OA} The overall architecture of SCNet.} 
\vspace{-0.1cm}
\end{figure*}

\begin{figure*}[htb]
\vspace{-0.1cm}
\center{\includegraphics[width=1\linewidth]  {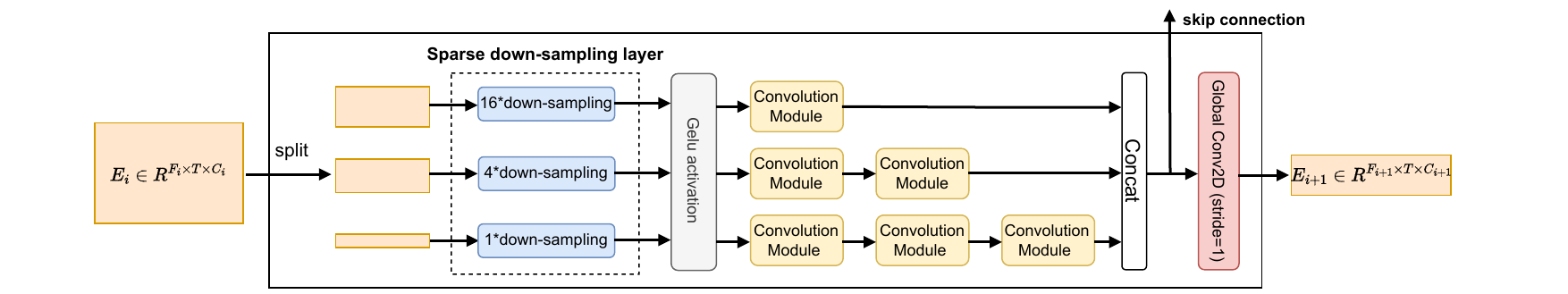}} 
\caption{\label{SD} The detailed architecture of sparse down-sampling block (SD block).} 
\vspace{-0.2cm}
\end{figure*}

\begin{figure*}[htb]
\vspace{-0.1cm}
\center{\includegraphics[width=0.9\linewidth]  {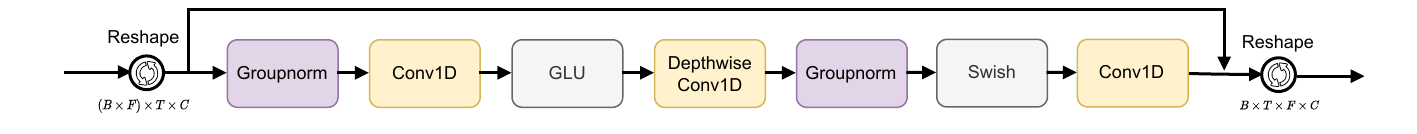}} 
\caption{\label{CM} The architecture of convolution module.} 
\vspace{-0.1cm}
\end{figure*}

\section{Method}
\subsection{Overall structure}
\label{sec:format}

Fig.\ref{OA} illustrates the entire workflow of SCNet. The original mixture is initially subjected to a Short-Time Fourier Transform (STFT) calculation:
\begin{equation}
y={\rm STFT}(x)
\end{equation}
For a given waveform $x$, the STFT operation converts the waveform into a complex spectrogram $y \in R^{F \times T \times(2\cdot2)}$, where $F$ and $T$ denote the frequency and time dimensions, respectively. The term $(2\cdot2)$ signifies the binaural channels and their respective real and imaginary components. The model takes $y$ as its input and is composed of three main parts: an audio encoder, a separation network, and an audio decoder, which is similar to Conv-TasNet \cite{luo2019conv}. The encoder aims to compress the spectrogram while enhancing its feature dimensions. The separation network then endeavors to learn sequence dependencies. Ultimately, the decoder reconstructs the spectrogram of each source.
In the following, the separation step is elaborated. %will be shown in detail.

\subsection{Encoder}
The encoder is designed to reduce the resolution of ultra-wideband audio, laying the groundwork for subsequent processing steps. For an input $y$, the encoder processes $y$ step by step through three sparse down-sampling blocks (SD blocks). Each traversal through a block modifies the frequency dimension's length from $F_{in}$ to $F_{in}\cdot R$, where $R$ represents the retention rate. Concurrently, the feature dimension undergoes adjustments, contingent upon the hyperparameters we establish.

SD blocks stand as the encoder's centerpiece. Fig.\ref{SD} shows its detailed architecture, incorporating a sparse down-sampling layer alongside a series of stacked convolution modules. Sparse down-sampling layer (SD layer) engineered with three parallel convolution layers to compress the frequency axis, the strides of these convolutions are 1, 4, and 16 respectively. 
Such a configuration indicates a tripartite division of the entire frequency band, applying a higher compression ratio in the high frequency part. In order to reduce the model complexity, we use the same frequency band division ratio in the down-sampling layers of all SD blocks. Additionally, SD layer contributes to an elevation in the feature dimension, enabling the model to capture more intricate details.
A GELU activation function \cite{hendrycks2016gaussian} ensues post the down-sampling layer.

The design of the convolution module is inspired by the Conformer \cite{gulati2020conformer}. As illustrated in Fig.\ref{CM}, we have opted for Groupnorm in lieu of Layernorm and Batchnorm as presented in the original design. In the low-frequency segment, we have incorporated additional convolutional blocks to enhance the granularity of our modeling.

On the whole, the encoder module prioritizes the preservation of low-frequency details.

%\subsection{Dual path module}
\subsection{Separation with dual-path module}
The encoder primarily focuses on modeling inter subband dependencies due to the pronounced differences in information density across various subbands. After compressing sufficiently, the information density becomes more uniform, shifting our attention to global information modeling.
To this end, we employ the dual-path RNN architecture from BSRNN. However, stacking multiple dual-path layers for the same sequence learning leads to limited performance due to diminishing marginal utility.
TFCNet \cite{tong2023tfcnet} addresses this by introducing a feature conversion module to project features into a new space. Inspired by this, we integrate the $torch.rfft$ and $torch.irfft$ functions \cite{paszke2019pytorch} between adjacent dual-path layers. Specifically, the output from odd-numbered dual-path layers is denoted as $Y_{i} \in R^{B \times F_{r} \times T \times C}(i=1,3,...)$, with $F_{r}$ signifying the compressed frequency dimension. The $rfft$ function is applied:
\begin{equation}
\begin{aligned}
\dot{Y_{i}}= {\rm RFFT}(Y_{i}[b,f ,:,c]) \in R^{B \times F_{r}\times (\frac{T}{2}+1) \times C}\quad\\
b=1,\cdot\cdot\cdot,B\quad f=1,\cdot\cdot\cdot,Fr\quad c=1,\cdot\cdot\cdot,C\quad \quad
\end{aligned}
\end{equation}
where $\dot{Y_{i}}$ is a complex tensor. This tensor is split into its real and imaginary components, which are then concatenated in the feature dimension, resulting in $X_{i+1} \in R^{B \times F_{r} \times (\frac{T}{2}+1) \times 2C}$ for the next layer. For the output of even-numbered layers, we adopt the inverse process of the aforementioned conversion.

\subsection{Decoder}
The decoder employs skip connections to integrate hierarchical features from the encoder, gradually reconstructing the separated spectrogram via the sparse up-sampling layer (SU layer). Within these skip connections, we incorporate a fusion layer. Initially, the two inputs are summed together, and the result is duplicated across the feature dimension. Subsequently, a 2-D convolution with a kernel size of 3 and a stride of 1 is applied, followed by a gated linear unit (GLU) \cite{dauphin2017language} layer:
\begin{equation}
GLU(a,b)= a \times sigmoid(b)
\end{equation}
where input is split in half along feature dimension to form $a$ and $b$. Given the prior replication steps, the GLU operation induces an effect akin to the swish function.

The SU layer acts as the counterpart to the SD layer, implemented by transposed convolution. There is only one difference. For the bottom SU layer, the output channel dimension is contingent on the count of sources set for separation, resulting in 4 channels designated for each source.

\subsection{Loss function}
Previous work generally considers waveform similarity as the training target. But this is not so closely related to the spectrogram. So we use the root mean squared error (RMSE) loss of complex-valued spectrogram as the loss funtion:
\begin{equation}
\mathcal{L}_{s}= \sqrt{(r-\widehat{r})^{2} + (i-\widehat{i})^{2}}
\end{equation}
where $\widehat{r}$ and $\widehat{r}$ represent the real and imaginary parts of the source spectrogram respectively. 

%\begin{equation}
%\mathcal{L}= \frac{1}{N} \sum_{w,h}\mathcal{L}_{s}
%\end{equation}

%where $w$ and $h$ denote the window and hop lengths in the STFT operation, respectively. $N$ represent the number of different settings. In the experiments, we used two settings:
%$(w,h)=(4096,1024),(8192,1024)$

\section{EXPERIMENTAL SETUP}
\subsection{Dataset}
\label{sec:pagestyle}

We validate the proposed algorithms on MUSDB18-HQ \cite{MUSDB18HQ}, the most
popular dataset to date to benchmark binaural music separation algorithms. It contain 150 full lengths music tracks of different genres along with their isolated drums, bass, vocals and others stems. In addition to this, we also use MoisesDB \cite{pereira2023moisesdb} as a supplementary dataset. MoisesDB is a comprehensive multitrack dataset of 240 previously unreleased songs by 47 artists, covering 12 advanced genres. 
In the experiment, we will see how much performance improvement will be achieved by using MoisesDB as additional training data.
Besides, the model's adaptability will also be tested on MoisesDB.
%as well as the model's adaptability when tested on Moisesdb. 

\begin{table*}[htbp]
\vspace{-0.3cm}
  \centering
  \caption{ Results of different sparsity rates on MUSDB18-HQ. The test environment for CPU RTF is Intel(R) Xeon(R) Platinum 8372HC CPU @ 3.40GHz, single-threaded. For comparison, the CPU RTF of HT Demucs is 1.38s.}
  \setlength{\tabcolsep}{4mm}{
    \begin{tabular}{cccccccccc}
    \toprule 
    \multirow{2}{*}{\textbf{GCR}} & \multicolumn{3}{c}{\textbf{SR}} & \multirow{2}{*}{\textbf{All}} & \multirow{2}{*}{\textbf{Drums}} & \multirow{2}{*}{\textbf{Bass}} & \multirow{2}{*}{\textbf{Other}} & \multirow{2}{*}{\textbf{Vocals}} & \multirow{2}{*}{\textbf{CPU RTF(s)}}\\
    \cmidrule(lr){2-4}
     &Low &Mid &High &  &  &  & &  & \\
    \midrule
    80\% &10\% &23.3\% &66.7\% & 7.66 & 9.26 & 7.6& 5.5& 8.24 &0.255\\
    75\% &15\% &25\% &60\% &8.52  &10.11  &8.57 &6.29 &9.11  &0.458\\
    \midrule
    \multirow{5}{*}{70\%}   &10\% &76.7\% &13.3\% &8.44 &10.01  &8.22  &6.16 &9.35 &0.659\\
       &12.5\% &64.2\% &23.3\% &8.58 &10.09  &8.37  &6.31 &9.54 &0.660\\
       &15\% &51.7\% &33.3\% &8.95  &10.43  &8.77 &6.63 &\textbf{9.93}  &0.663\\
       &17.5\%\dag &39.2\% &43.3\% &\textbf{9.0}  &\textbf{10.51}  &\textbf{8.82} &\textbf{6.76} &9.89  &0.669\\
       &20\% &26.7\% &53.3\% &8.87  &10.42  &8.70 &6.64 &9.71  &0.676\\
    \bottomrule
    \end{tabular}}
  \label{sparceresult}
\vspace{-0.3cm}
\end{table*}

\subsection{Model and training configurations}
During training, we split the soundtrack 
into overlapped segments of duration 11 seconds. The time interval between two adjacent segments is 1 second. Additionally, we adopt the data augmentation methods, remix and scale, following the ones used in Demucs \cite{defossez2019demucs}. Remix achieves data simulation by dynamically mixing tracks of different songs during training, while scale enhances model robustness by randomly adjusting the volume of tracks.

The STFT window size is 92ms (4096-point FFT) with a hop size of 23ms. Thus, the spectrogram will have 2049 frequency bins. We do not use a window function in the STFT operation, as this has no effect on the results. Three sparse down-sampling layers (SD layers) sequentially increase features to 32, 64, and 128 dimensions.
The hidden layer size of the convolutional module is one-fourth of the input size, with the three convolution layers in its middle having kernel sizes of 3, 3, and 1, respectively. We stack 6 dual-path RNN layers in the separation network. At odd layers, the number of hidden units is 128 in BiLSTM, which goes up to 256 in even-numbered layers.

All our experiments are conducted on 8 Nvidia V100 GPUs. When training solely on the MUSDB18-HQ dataset, the model is trained for 130 epochs with the Adam \cite{kingma2014adam} optimizer with an initial learning rate of 5$e$-4 and batch size of 4 for each GPU. Nevertheless, we adjust the learning rate to 3$e$-4 when introducing additional data to mitigate potential gradient explosion.

\subsection{Evaluation metrics}
In the experiments, we utilize Signal-to-Distortion Ratio (SDR) \cite{vincent2006performance}, as computed by museval \cite{stoter20182018} as the primary evaluation metric. Furthermore, Real-Time Factor (RTF) is also used to assess the model's efficiency during testing. RTF is defined as the ratio of the time required to process a fixed input to the duration of that input.

% \begin{table*}[htbp]
% \vspace{-0.3cm}
%   \centering
%   \caption{ Results of different sparsity rates on MUSDB18-HQ. The test environment for CPU RTF is Intel(R) Xeon(R) Platinum 8372HC CPU @ 3.40GHz, single-threaded. For comparison, the CPU RTF of HT Demucs is 1.38s.}
%   \setlength{\tabcolsep}{4mm}{
%     \begin{tabular}{cccccccccc}
%     \toprule 
%     & \multicolumn{3}{c}{\textbf{SR}}  \\
%     \cmidrule(lr){2-4}
%     \textbf{GCR} &Low &Mid &High & \textbf{ALL} & \textbf{Drums} & \textbf{Bass} &\textbf{Other} & \textbf{Vocals} &\textbf{CPU RTF(s)}\\
%     \midrule
%     80\% &10\% &23.3\% &66.7\% & 7.66 & 9.26 & 7.6& 5.5& 8.24 &0.34\\
%     75\% &15\% &25\% &60\% &8.52  &10.11  &8.57 &6.29 &9.11  &0.61\\
%     \midrule
%     \multirow{5}{*}{70\%}   &10\% &76.7\% &13.3\% &8.44 &10.01  &8.22  &6.16 &9.35 &0.880\\
%        &12.5\% &64.2\% &23.3\% &8.58 &10.09  &8.37  &6.31 &9.54 &0.882\\
%        &15\% &51.7\% &33.3\% &8.95  &10.43  &8.77 &6.63 &\textbf{9.93}  &0.884\\
%        &17.5\% &39.2\% &43.3\% &\textbf{9.0}  &\textbf{10.51}  &\textbf{8.82} &\textbf{6.76} &9.89  &0.891\\
%        &20\% &26.7\% &53.3\% &8.87  &10.42  &8.70 &6.64 &9.71  &0.901\\
%     \bottomrule
%     \end{tabular}}
%   \label{sparceresult}
% \end{table*}

\section{RESULTS}
\subsection{Effect of sparsity rate}
Table \ref{sparceresult} shows the influence of different sparsity rates on performance. The term global compression ratio (GCR) denotes the reduction percentage of the frequency dimension following each SD block. Sparsity rate (SR) indicates the division strategy of the three frequency bands. As the results show, a reduced GCR will significantly improve performance, however, it concurrently increases computational demands. To manage computational expenses, we refrain from testing even lower GCR values.
While SR exerts a minimal influence on efficiency, its correct selection is paramount. The results suggest that peak performance is reached when the share of low-frequency content approximates 17.5\%. If it falls below this range, it might cause the loss of essential information. Conversely, exceeding this range can lead to a reduced range of the mid-frequency band, thereby impacting modeling effectiveness. Based on our experimental findings, we have selected a configuration with GCR set at 70\% and a low-frequency proportion of 17.5\% for our final model, as it delivers the optimal average SDR.

\subsection{Results on MUSDB18-HQ}
Table \ref{musdbresult} compares the proposed SCNet with other previous works on the MUSDB18-HQ dataset. SCNet achieves 9.0 dB mean SDR on the test set without extra training data. Remarkably, our model marks a considerable advancement in the SDR metrics for drums and bass. Simultaneously, it keeps pace with leading-edge models in SDR metrics for vocals and other. We also utilize an additional dataset of 235 four-track songs provided by MoisesDB for training. By augmenting the training data, we observe improvements in the performance across all tracks. Furthermore, it's worth noting that SCNet boasts a notably streamlined architecture with just 10.08M parameters, which is a mere quarter of the parameter count in HT Demucs. Attempting to stack more parameters in the large version, we doubled the channel dimension and conducted training using PyTorch's mixed precision approach. This configuration resulted in a total of 41.2M parameters.

\begin{table}[htbp]
\vspace{-0.45cm}
  \centering
  \caption{Comparison with other methods on MUSDB18-HQ (results with a * are tested on the non HQ version). ``Extra" represents the quantity of additional data used during training. BSRNN only utilizes mixtures, while the others employ multi-track songs.}
  \setlength{\tabcolsep}{0.9mm}{
    \begin{tabular}{ccccccc}
    \toprule
    \textbf{Model} & \textbf{Extra}& \textbf{All} & \textbf{drums} & \textbf{bass} &\textbf{other} & \textbf{vocals} \\
    \midrule
    HT Demucs \cite{rouard2023hybrid} &No & 7.52 & 7.94 & 8.48  & 5.72 &7.93\\
    Hybrid Demucs \cite{defossez2021hybrid} &No &7.64 &8.12 &8.43 &5.65 &8.35\\
    ResUNet* \cite{kong2021decoupling} &No &6.73 &6.62 &6.04 &5.29 &8.98\\
    BSRNN \cite{luo2023music}    &No &8.24 &9.01 &7.22 &6.70 &10.01\\
    \midrule
    SCNet  &No  &9.00 &10.51 &8.82  &6.76 &9.89\\
    SCNet-large  &No  &\textbf{9.69} &\textbf{10.98} &\textbf{9.49}  &\textbf{7.44} &\textbf{10.86}\\
    \midrule
    \midrule
    Spleeter* \cite{hennequin2020spleeter}  &2500 & 5.91 & 6.71 & 5.51  & 4.55 &6.86\\
    D3Net \cite{takahashi2020d3net}  &1500  &6.68 &7.36 &6.20  &5.37 &7.80\\
    Hybrid Demucs \cite{defossez2021hybrid} &800 &8.34 &9.31 &9.13 &6.18 &8.75\\
    HT Demucs \cite{rouard2023hybrid} &800  &9.00 &10.08 &\textbf{10.39} &6.32 &9.20\\
    BSRNN \cite{luo2023music} &1750 &8.97 &10.15 &8.16 &7.08 &10.47\\
    \midrule
    SCNet  &235  &9.25 &10.78 &9.21  &6.84 &10.17\\
    SCNet-large  &235  &\textbf{9.92} &\textbf{11.23} &9.86  &\textbf{7.51} &\textbf{11.1}\\
    \bottomrule
    \end{tabular}}
  \label{musdbresult}
\vspace{-0.5cm}
\end{table}

\subsection{Results of generalizability test on MoisesDB}
We further assess the generalizability of our approach in Table \ref{generalizability}. SCNet is trained on the MUSDB18-HQ and evaluated on the MoisesDB without fine-tuning. As a point of comparison, HT Demucs was trained with an additional 800 songs. The results demonstrate that SCNet exhibits superior generalization even with limited data.

\begin{table}[htbp]
\vspace{-0.45cm}
  \centering
  \caption{Comparison of generalizability with HT Demucs.}
  \setlength{\tabcolsep}{1.3mm}{
    \begin{tabular}{ccccccc}
    \toprule
    \textbf{Model} & \textbf{Extra}& \textbf{All} & \textbf{drums} & \textbf{bass} &\textbf{other} & \textbf{vocals} \\
    \midrule
    HT Demucs \cite{rouard2023hybrid} &800 & 9.91 & 10.94 & 11.64  & 7.00 &10.05\\
    \midrule
    SCNet  &No  &\textbf{10.33} &\textbf{11.40} &\textbf{11.65}  &\textbf{7.54} &\textbf{10.74}\\

    \bottomrule
    \end{tabular}}
  \label{generalizability}
\vspace{-0.5cm}
\end{table}

\subsection{Ablation experiments}
To further validate the effectiveness of sparse compression, we conduct ablation experiments. Presented in Table \ref{Ablation}, the control group utilizes the same frequency segmentation method but applies 4*down-sampling to each subband, retaining a GCR of 75\%. The results indicate that sparse compression is crucial for achieving our desired outcomes.

\begin{table}[htbp]
\vspace{-0.45cm}
  \centering
  \caption{Results of ablation experiments.}
  \setlength{\tabcolsep}{1.3mm}{
    \begin{tabular}{cccccc}
    \toprule
    \textbf{Model} & \textbf{All} & \textbf{drums} & \textbf{bass} &\textbf{other} & \textbf{vocals} \\
    \midrule
    SCNet (GCR=75\%)  & \textbf{8.52} & \textbf{10.11} & \textbf{8.57}  & \textbf{6.29} &\textbf{9.11}\\
    \midrule
    w/o sparse compression  &6.20 &7.34 &5.48  &4.54  &7.45 \\

    \bottomrule
    \end{tabular}}
  \label{Ablation}
\vspace{-0.5cm}
\end{table}

\section{Conclusion}
\label{sec:majhead}
In this paper, we propose an innovative frequency-domain network architecture called SCNet, which explicitly splits the spectrogram of the mixture into three subbands and introduces a sparsity-based encoder to model different subbands. By applying a higher compression ratio to the mid- and high-frequency subbands, we enhance the overall information density. This heightened density facilitates a more efficient utilization of the separation network. Experiment results show that, with lower computational consumption, SCNet can surpass the performance of existing state-of-the-art music source separation methods on the MUSDB18-HQ dataset, whether using extra data or not. In the future, we will study better sub-band splitting methods and further improve the separation module.
% such as using other architecture like transformers.

\section{ACKNOWLEDGEMENT}
\label{sec:majhead}
This work is supported by Shenzhen Science and Technology Program (WDZC20200818121348001, JCYJ20220818101014030), the Major Key Project of PCL (PCL2021A06, PCL2022D01).

\bibliographystyle{IEEEbib}
\bibliography{strings,refs}

\end{document}